\documentclass[12pt,a4paper,final]{iopart}

\usepackage{iopams}  
\usepackage{graphicx}
\usepackage[breaklinks=true,colorlinks=true,linkcolor=blue,urlcolor=blue,citecolor=blue]{hyperref}

\usepackage{tikz}
\newcommand*\circled[1]{\tikz[baseline=(char.base)]{
            \node[shape=circle,draw,inner sep=2pt] (char) {#1};}}
            
\begin{document}

\title{Energy harvesting via an adaptive bistable potential}

\author{Ashkan Haji Hosseinloo}
\address{Massachusetts Institute of Technology, 77 Massachusetts Avenue,
Cambridge MA}
\ead{ashkanhh@mit.edu}

\author{Konstantin Turitsyn}
\address{Massachusetts Institute of Technology, 77 Massachusetts Avenue,
Cambridge MA}
\ead{turitsyn@mit.edu}

\begin{abstract}
Narrow bandwidth and easy detuning, inefficiency in broadband and non-stationary excitations, and difficulties in matching linear harvester's resonance frequency to low-frequency excitations at small scales, have convinced the researchers to investigate the nonlinear, in particular the bistable energy harvesters in the recent years. However, the bistable harvesters suffer from co-existing low and high energy orbits, and sensitivity to initial conditions, and have been recently proven inefficient when subjected to many real-world random and non-stationary excitations. Here, we propose a novel buy-low-sell-high strategy that can significantly improve the harvester's efficiency at low-frequencies in a much more robust fashion. This strategy could be realized by a passive adaptive bistable system. Simulation results confirm high efficiency of the adaptive bistable system following a buy-low-sell-high logic when subjected to harmonic and random non-stationary walking excitations compared to its conventional bistable and linear counterparts.  

\end{abstract}

\vspace{2pc}
\noindent{\it Keywords}: energy harvesting, bistable potential, adaptive potential, buy-low-sell-high strategy

\section{Introduction}

Short life span, miniaturization and scalability difficulties, replacement and maintenance issues, and relatively very low pace of energy density improvement of conventional batteries\cite{anton2007review} have convinced many researchers and scientists to consider energy harvesters as potential replacement for batteries in many applications. In particular, vibratory energy harvesters have captured  enormous attention in the last decade due to universality and abundant availability of vibratory energy sources.

Linear harvesters exploiting resonance phenomena suffer from narrow bandwidth of efficient harvesting. The narrow resonance bandwidth renders linear harvesters very inefficient when subjected to non-stationary excitation where excitation characteristics e.g. dominant frequency, change over time, or when the harvester is exposed to broadband random vibration where the excitation power is spread over a wide frequency range. To overcome this issue different techniques such as resonance tuning, multi-modal energy harvesting, frequency up-conversion, and more recently purposeful inclusion of nonlinearity have been suggested \cite{Tang2010,daqaq2014role}. Among these different techniques, deliberate introduction of nonlinearity in particular, bistable nonlinearity have been the focus of the research in vibration energy harvesting since 2009. However, recent studies have revealed that monostable and bistable nonlinear harvesters do not always outperform their linear counterparts.

One of the main issues with the bistable harvester when subjected to harmonic excitation is non-uniqueness of the solution and co-existing low-energy and high-energy orbits at a given excitation frequency and amplitude \cite{erturk2009piezomagnetoelastic,mann2010investigations,stanton2010nonlinear,erturk2011broadband}. In fact, for a monostable nonlinear harvester the probability of converging to the low-energy orbit is higher than that of the high-energy orbit \cite{quinn2011comparing}. Also, Masana and Daqaq \cite{masana2011relative} showed that for a given excitation level, bistable harvester's performance is very sensitive to the potential shape (shallow versus deep wells).

Performance of the bistable harvester is further diminished when it is subjected to random excitation. Daqaq \cite{daqaq2011transduction} showed that for an inductive energy harvester with negligible inductance, bistability (in general any stiffness nonlinearity) does not provide any improvement over the linear one when excited by white noise. Cottone et al. \cite{cottone2009nonlinear} and Daqaq \cite{daqaq2012intentional} showed that when driven by white noise, a necessary condition for the bistable harvester to outperform its linear counterpart is to have a small ratio of mechanical to electrical time constants. They along with other researchers \cite{litak2010magnetopiezoelastic,halvorsen2013fundamental,zhao2013stochastic} showed that for a given noise intensity, the output power highly depends on the shape of the bistable potential. Zhao and Erturk \cite{zhao2013stochastic} showed that the bistable harvester could outperform its linear counterpart only in a narrow region where noise intensity is slightly above the threshold of interwell oscillations.

The bistable harvester becomes even less efficient and less robust when it is excited by more realistic and real-world random vibrations (not white noise). Using real vibration measurements (human walking motion and bridge vibration) in simulations of idealised energy harvesters Green et al. \cite{green2013energy} showed that, although the benefits of deliberately inducing dynamic nonlinearities into such devices have been shown for the case of Gaussian white noise excitations, the same benefits could not be realised for the real excitation conditions.

In this paper, we propose an adaptive bistable harvester that is more robust to changes in input excitation parameters and works more efficiently under both harmonic and random excitations when compared to its conventional linear and bistable counterparts. In the proposed harvester, the potential barrier changes adaptively following a buy-low-sell-high strategy \cite{hosseinloo2014fundamental}.

\section{Adaptive bistable harvester}

In this study, we consider both capacitive and inductive harvesters (with single-degree-of-freedom in the mechanical domain) with an adaptive bistable potential. Here, the adaptive bistable potential refers to a potential where the potential shape, in particular the potential barrier height could change according to a logic in an adaptive fashion. Adaptive bistability could be realized in different ways that will briefly be discussed later in the paper. But first we need to find the logic according to which the bistability changes adaptively. The strategy to maximize the harvested energy for a general and ideal harvester was derived and discussed in detail by the authors in \cite{hosseinloo2014fundamental}. However, for the sake of readability and its application to the bistable system, its key concepts are discussed next. 

\subsection{Adaptive bistability logic: Buy-Low-Sell-High}

To find the logic, we consider a model of a single-degree-of-freedom ideal energy harvester characterized by the mass $m$ and the displacement $x(t)$ that is subjected to the energy harvesting force $f(t)$ and exogenous excitation force $F(t)$. Then, the equation of the motion will simply be:
\begin{equation}
 m\ddot{x}(t)=F(t)+f(t).
 \label{Eq:1}
\end{equation}
Here we assume that the ideal harvesting force can harvest all the energy that flows to the system (there is no accumulation of energy in the system in long-term). Hence, maximizing the harvested energy will be equivalent to maximizing the energy flow to the system. In other words, we want to maximize:
\begin{equation}
   E_{\mathrm{max}} = \max_{x(t)} \int\mathrm{d}t \, F(t)\dot{x}(t),
   \label{Eq:2}
\end{equation}
over admissible trajectories of $x(t)$. It is easy to show that this integral is unbounded if $x(t)$ is unconstrained. Indeed, the trajectory defined by a simple relation $\dot{x}(t)=\lambda F(t)$ that can be realized with the harvesting force $f = m\lambda\dot{F} - F$ results in the harvesting rate of $\lambda F^2$ that can be made arbitrarily large by increasing the mobility constant $\lambda$. This trivial observation illustrates that the question of fundamental limits is only well posed for the model that incorporate some technological or physical constraints. This is a general observation that applies to most of the known fundamental limits. For example, Carnot cycle limits the efficiency of cycles with bounded working fluid temperature, and Shannon capacity defines the limits for signals with bounded amplitudes and bandwidth.

As a common constraint to vibratory energy harvesters, we constrain the harvester displacement in a symmetric fashion i.e. $|x(t)|\leq x_{\mathrm{max}}$, where $x_{\mathrm{max}}$ is the displacement limit. Rewriting Eq.\ref{Eq:2} as $-\int\mathrm{d}t \, \dot{F}(t) x(t)$, it could be seen by inspection that the integral is maximized by the optimal trajectory:
\begin{equation}
 x_*(t)=-x_{\mathrm{max}} \, \mathrm{sign}\left[\dot{F}(t)\right].
 \label{Eq:3}
\end{equation}
This optimal trajectory is indeed realizable by the ideal harvesting force of $f(t)=m\ddot{x}_*(t)-F(t)$. The interpretation of Eq.\ref{Eq:3} is easy; it says when $F(t)$ is increasing, $x(t)$ should be kept at its lowermost limit, and vice versa, when $F(t)$ is decreasing, $x(t)$ should be kept at its uppermost limit. Thus, the transitions between displacement limits occur when sign of $\dot{F}(t)$ is changing i.e. at extremums of $F(t)$.
In other words, in this logic, the harvester mass is kept at its lowest position ($-x_{\mathrm{max}}$) until the excitation force $F(t)$ reaches its maximum when the mass should then be pushed to its highest position ($x_{\mathrm{max}}$) (either by the excitation force, or by the harvesting force if the local maximum of the excitation force is still negative or not big enough to push the mass to the highest position limit\footnote{It should be noted even though the harvesting force is injecting energy to the system in this case during a short period, the net amount of harvested energy will be positive at the end. This is because injection of the energy by the harvesting force will pay off when the next excitation force minimum is reached.}). Similar dynamics occur in the reverse direction and this strategy continues in the same fashion at every extremum of the excitation force $F(t)$.

If the harvester is incapable of injecting energy to the system (passive-only harvester), the harvested mass should traverse between the limits ($\pm x_{\mathrm{max}}$) by the excitation force $F(t)$ only. In this case, the logic is slightly modified; the harvester mass should be kept at its lowest (highest) displacement limit till the largest maximum (most-negative minimum) of the excitation force is reached. Only then, the harvester mass is pushed from one displacement limit to the other. This logic is very similar to the well-known buy-low-sell-high strategy in stock market; hence, we call this logic a Buy-Low-Sell-High (BLSH) strategy hereafter.

\begin{figure}
 \centering
 \includegraphics[width = 0.8 \linewidth]{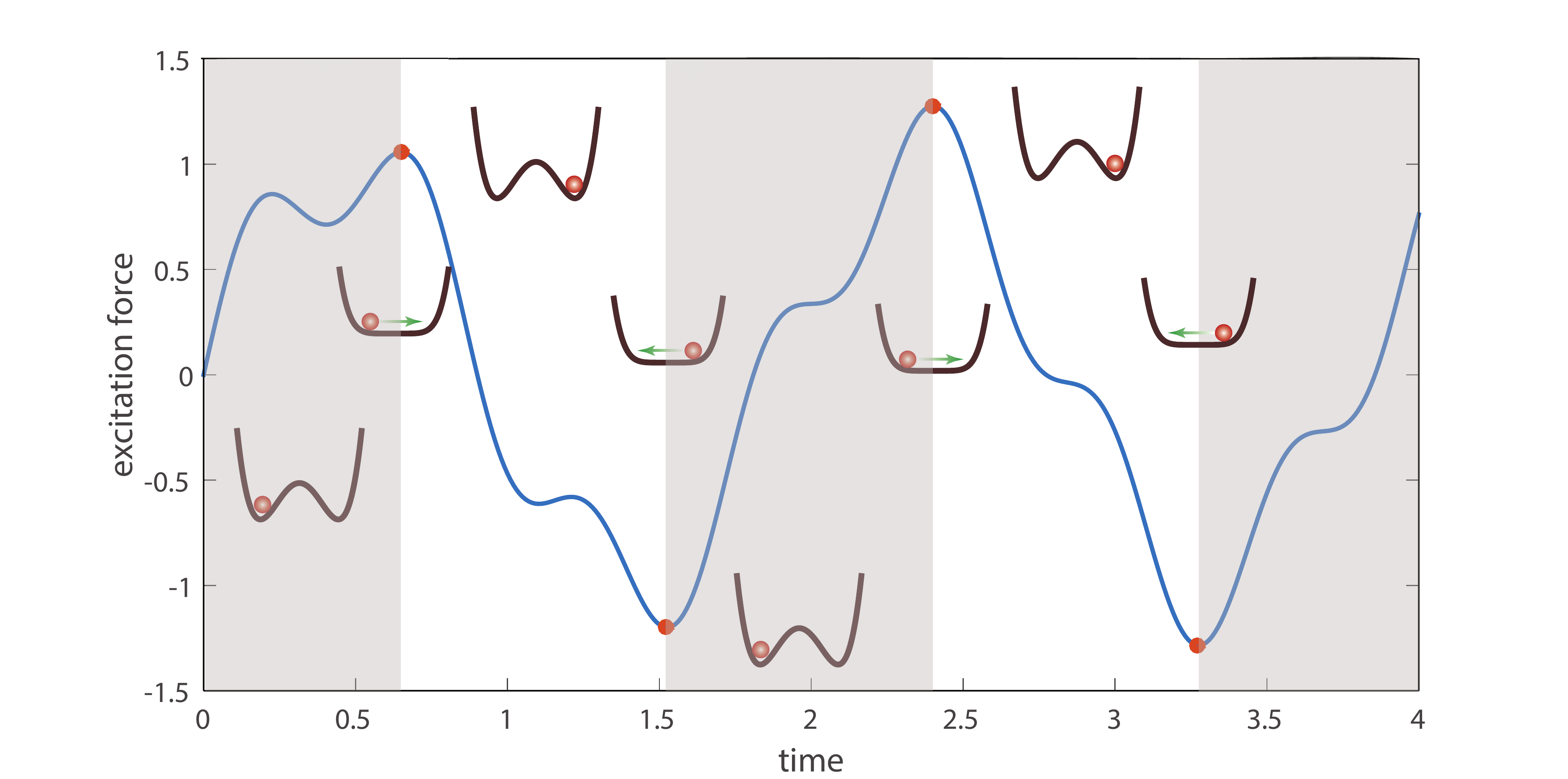}
 \caption{Passive BLSH strategy realized by an adaptive bistable potential for an arbitrary excitation input. The transition from one displacement limit to the other is highlighted by the background colour change in the figure.}
 \label{Fig1}
\end{figure}

Now the question is how to implement this logic. The BLSH strategy could be realized by an adaptive bistable potential. In essence, the passive BLSH strategy keeps the harvester mass at one end ($\pm x_{\mathrm{max}}$) before letting it go to the other end according to its logic. A bistable potential with stable points at $\pm x_{\mathrm{max}}$ and adaptive potential barrier could do this. To realize the BLSH logic, the potential barrier should be large enough to confine the harvester mass in one well ($x_{\mathrm{max}}$ or $-x_{\mathrm{max}}$). Then, when, according to the logic, the harvester mass should travesre to the other end the potential barrier should vanish. This logic is schematically shown in Fig.\ref{Fig1}.

\subsection{Mathematical modeling}

The harvester is modeled as a lumped-parameter mechanical oscillator coupled to a simple electrical circuit via an electromechanical coupling mechanism. The formulation here is generic and could be applied to both capacitive (e.g. piezoelectric) and inductive (e.g. electromagnetic) transduction mechanisms. The nondimensionalized governing dynamic equations could be written as \cite{daqaq2014role}:

\begin{eqnarray}
\ddot{x}+2\zeta \dot{x}+\frac{\partial U(x,t)}{\partial x}+\kappa^2 y=-\ddot{x}_b \nonumber \\
\dot{y}+\alpha y=\dot{x}.
\label{Eq:4}
\end{eqnarray}
 
In the above equations, $x$ is the oscillator's displacement relative to base displacement ($x_b$). Linear mechanical damping is characterized by the damping ratio $\zeta$, and $\kappa$ denotes the linear electromechanical coupling coefficient. $y$ represents the electric quantity that would be voltage or current in capacitive or inductive transduction mechanisms, respectively and $\alpha$ is the ratio of the mechanical to electrical time constants. The adaptive bistable potential is denoted by $U(x,t)$ and overdot denotes differentiation with respect to nondimensional time. All parameters and variables are nondimensional.

Two common techniques to realize bistability are buckling phenomenon and magnets (to create negative stiffness) in addition to the positive mechanical stiffness. When using magnetic field to realize bistability, if permanent magnets are replaced by electromagnets \cite{ouellette2014broadband} (thus having a controllable magnetic field) one can change the potential shape; hence, create an adaptive bistability. A passive bistable potential admits a quartic form \cite{tang2012improving}, and when made adaptive, we model it as:

\begin{equation}
U(x,t)=\frac{1}{2}(1+\delta(t)r_k)x^2-\frac{1}{4}\delta(t)(1+r_k)\frac{x^4}{x_s^2},
\label{Eq:5}
\end{equation}
where $r_k<-1$ is strength of the negative stiffness of the magnetic field relative to the linear mechanical one. $x_s$ denotes the nondimensional stable position of the bistable potential, and $\delta(t)$ is a Heaviside function which switches between 1 and 0 according to the BLSH logic. $\delta(t)$ is always equal to unity except when we want the harvester mass traverse from one end to the other (according to the BLSH strategy) which then is set to zero.

\begin{figure}
 \centering
 \includegraphics[width = 0.9 \linewidth]{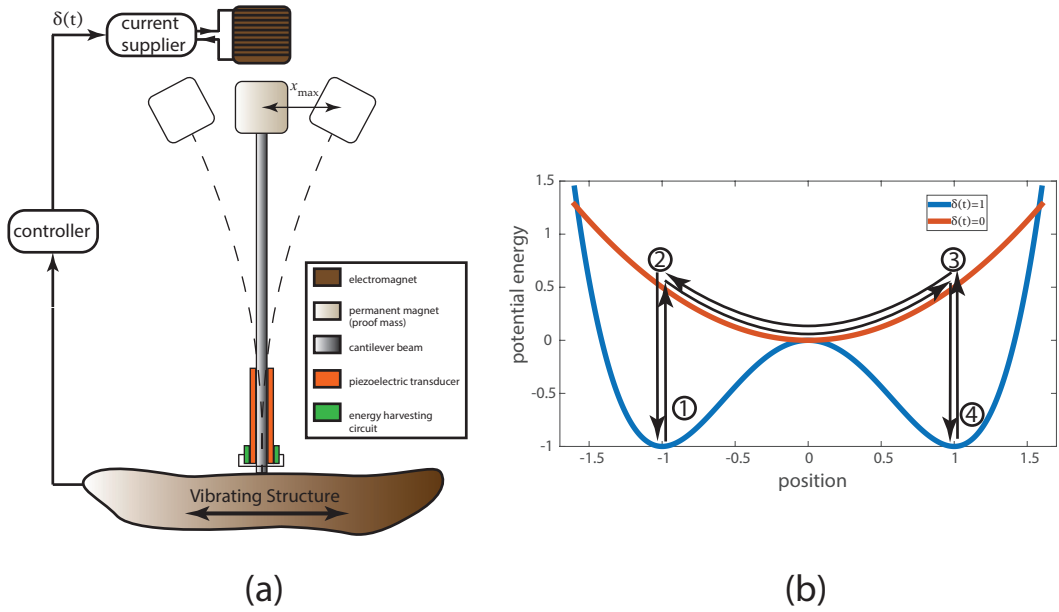}
 \caption{Energy harvesting with adaptive bistability (a) schematics of a cantilever energy harvester with piezoelectric transduction mechanism equipped with adaptive bistability (b) change in harvester's potential function to realize the BLSH logic and the sequence of the harvester mass trajectory on admissible potential curves following the logic}
  \label{Fig:2}
\end{figure}

Figure\ref{Fig:2}(a) depicts an energy harvester with piezoelectric (capacitive) transduction mechanism equipped with adaptive bistable potential. The adaptive bistability is realized by an electromagnet and a permanent magnet (the proof mass). An On/Off controller is used to implement the BLSH logic. The controller senses the excitation and then according to the BLSH strategy sends a signal to the current supplier to supply an appropriate current ($\delta(t)=1$) or to shut down the current supply ($\delta(t)=0$).

Figure\ref{Fig:2}(b) shows how the potential shape changes by the controller signal $\delta(t)$, and graphically depicts the sequence of the harvester mass trajectory following BLSH logic on admissible potential curves\footnote{In fact when the magnetic potential is added to the system, the whole bistable potential curve should be shifted above the quadratic mechanical potential curve. This does not show up here as we have dropped a constant term in Eq.\ref{Eq:5}. However this does not affect the dynamics of the system.}. It should be noted with this type of implementation (Eq.\ref{Eq:5} and Fig.\ref{Fig:2}(a)) the adaptive bistable system following BLSH logic will not be passive for all time. For instance when the harvester mass is moved $\circled{1}\rightarrow\circled{2}$ ($\circled{4}\rightarrow\circled{3}$) a positive amount of energy is added to the system because of the way the potential shape is changed. However, in the transition right before the one that adds energy, i.e. in $\circled{2}\rightarrow\circled{1}$ ($\circled{3}\rightarrow\circled{4}$) the same amount of energy is taken out of the system; hence, the net energy injected to the system by this type of implementation is zero in half a cycle (if not zero for all time) where cycle is referred to transitions from $-x_{\mathrm{max}}$ to $+x_{\mathrm{max}}$ and then back again to $-x_{\mathrm{max}}$. In order to have a passive system for all time, one should come up with a bistable mechanism whose potential barrier could be deepened without changing the potential energy level of its stable points e.g. like a latching mechanism. This is not the case with the current techniques for bistability realization (buckling and magnetic field).
\section{Results and discussion}
In this section, simulation results with harmonic and experimental random excitations for adaptive bistable harvester is presented and compared with linear and conventional bistable harvesters. For a fair comparison, all harvesters are subjected to the same displacement limits. To this end, we first optimize the bistable system with respect to its potential shape for given excitation input. Then the maximum displacement of the optimum bistable harvester is set as the maximum displacement limit for the linear and adaptive bistable systems. This approach greatly favors the conventional bistable system when it comes to comparison.

\subsection{harmonic excitation}
\begin{figure}
 \centering
 \includegraphics[width = \linewidth]{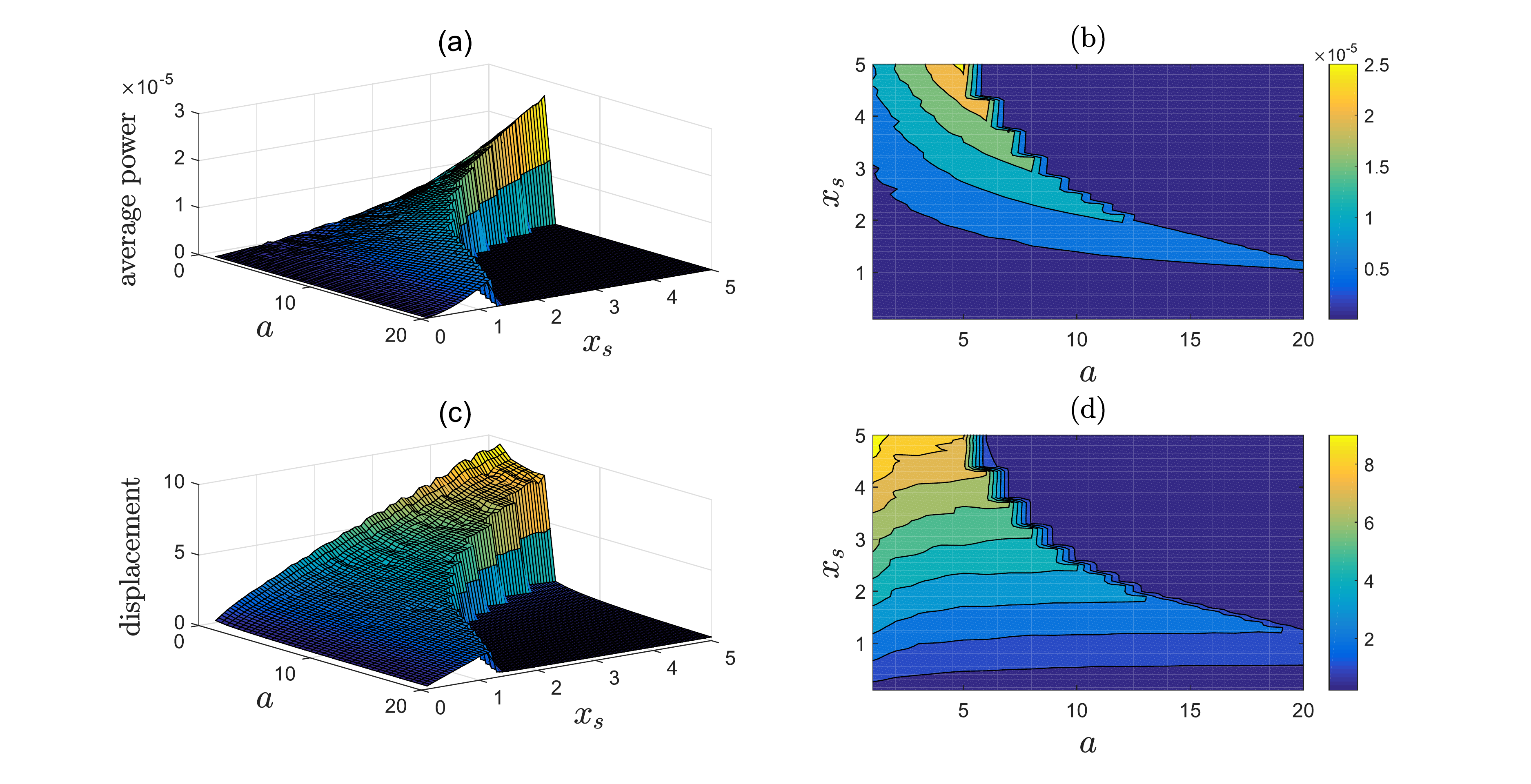}
 \caption{Energy harvesting with conventional bistable system. (a) and (b) show surface and contour plots of average harvested power in terms of system parameters $a$ and $x_s$. (c) and (d) show surface and contour plots of harvester displacement amplitude in terms of system parameters $a$ and $x_s$. The other parameters are set as $F_0=10$, $w=0.05$, $\zeta=0.01$, $\kappa=5$, and $\alpha=1000$.}
  \label{Fig:3}
\end{figure}

The potential function considered here for the bistable system is the same as the one used for the adaptive bistable harvester with a small change in the parameter notation ($1+r_k\rightarrow -a$). The potential used is of the form $U(x)=-\frac{1}{2}ax^2+\frac{1}{4}a\frac{x^4}{x_s^2}$ where $a>0$. Fig.\ref{Fig:3} shows the average power and displacement amplitude of the bistable system when subjected to harmonic excitation of the form $-\ddot{x}_b=F_0 \sin(\omega t)$. This paper intends to target mainly the low-frequency excitation where the linear harvesters fail to work efficiently; hence, the nondimensional excitation frequency used here is set to $\omega=0.05$. The average power is calculated by $\frac{1}{t}\int_0^t y^2(t) \mathrm{d}t$ for a long simulation time $t$. One should note that this expression gives the normalized nondimensional average power. The dimensional instantaneous power is equal to $(m \omega_n^3 l_c^2)\alpha \kappa^2 y^2$ where $m$, $\omega_n$, and $l_c$ are the harvester mass, time-scaling frequency, and length scale, respectively. Hence, the average power used here is nondimensionalized by $m \omega_n^3 l_c^2$, and further normalized by $\alpha \kappa^2$\footnote{Since we are not optimizing the power with respect to $\alpha$ and $\kappa$ it is fine to normalize the power by $\alpha\kappa^2$.}.

\begin{figure}
 \centering
 \includegraphics[width = 0.8 \linewidth]{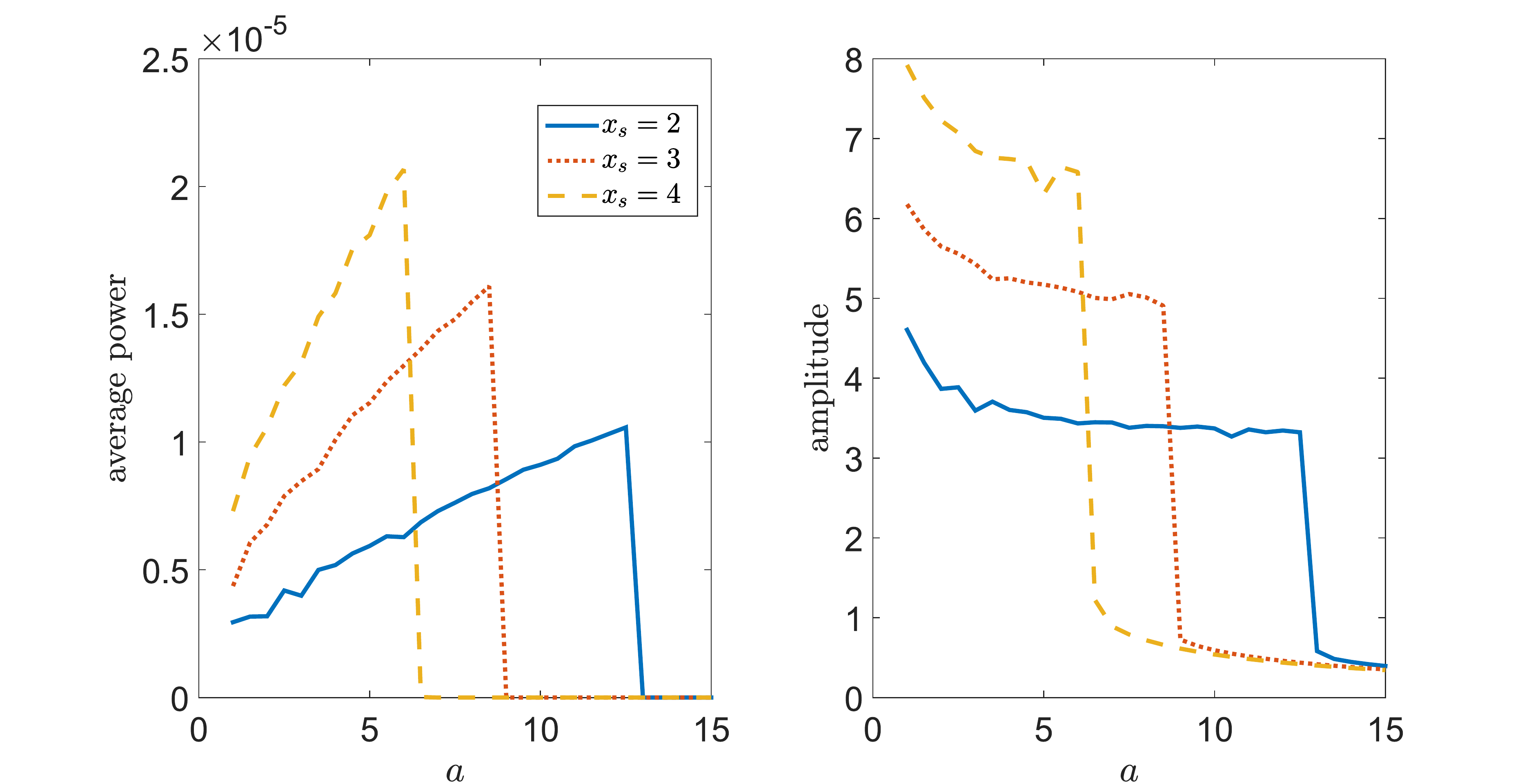}
 \caption{Average harvested power (on the left) and harvester displacement amplitude (on the right) of the conventional bistable energy harvester as a function of the potential parameter $a$ for three different values of the parameter $x_s=2,3,4$. }
  \label{Fig:4}
\end{figure}

It could be seen from Fig.\ref{Fig:3} that the average power increases monotonically with $a$ and $x_s$ up to a maximum and then drops sharply. This is where the interwell oscillation turns into intrawell oscillation (potential barrier linearly increases with $a$ and $x_s^2$). A drastic decrease in the amplitude of oscillation verifies this. It should be noted that for values below the optimum value of $a$ (for a given $x_s$), the system is still in interwell motion; however, the power monotonically decreases as $a$ is decreased from its optimum value. This could be seen more clearly in Fig.\ref{Fig:4}. This suggests the robustness issues with the conventional bistable system, that is, the harvester works efficiently only when the potential barrier is slightly below its critical value when it triggers the interwell oscillation which agrees with Zhao and Etrurk's claim \cite{zhao2013stochastic}.

Next, we compare the performance of the adaptive bistable harvester with that of an optimized conventional bistable and linear harvesters when they are subjected to harmonic excitation. To this end, we first optimize the parameters of the bistable system for given excitation input and displacement limits. The same harmonic excitation used in Figs. \ref{Fig:3} and \ref{Fig:4} is considered here ($F_0=10$ and $\omega=0.05$). According to Figs.\ref{Fig:3} and \ref{Fig:4} the optimal parameters corresponding to maximum displacement of 3.4 are $x_s=2$ and $a=12$. For a fair comparison the parameters of the adaptive bistable and linear harvesters are set such that their maximum displacements do not exceed this value ($r_k=-300$ and $x_s=2.8$ for the adaptive bistable, and natural frequency of $\sqrt{3}$ for the linear harvester).

\begin{figure}
 \centering
 \includegraphics[width = 0.8 \linewidth]{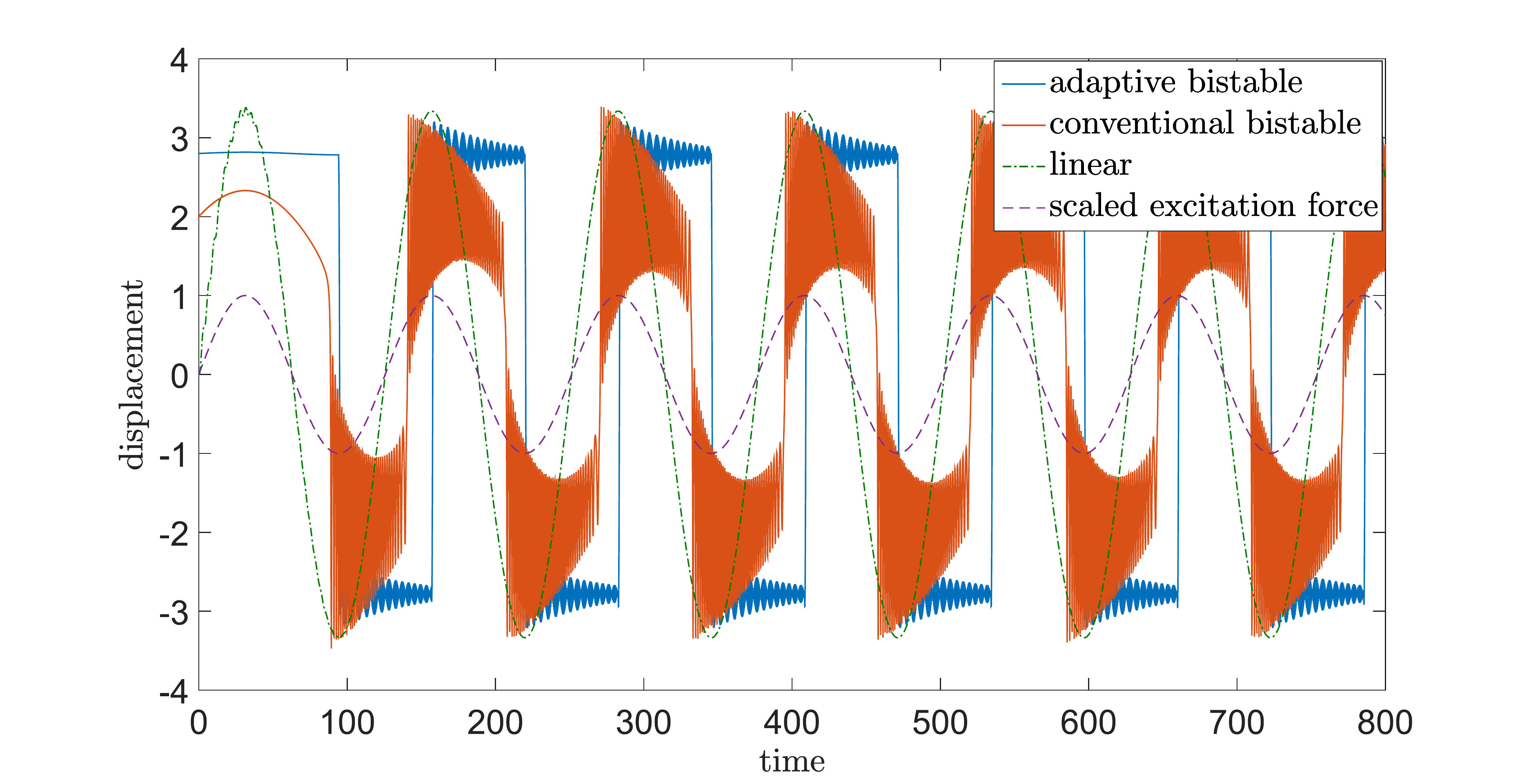}
 \caption{Displacement time histories of linear, conventional bistable, and adaptive bistable energy harvesters subjected to harmonic excitation with excitation amplitude $F_0$=10, and frequency $\omega$=0.05.}
  \label{Fig:5}
\end{figure}

\begin{figure}
 \centering
 \includegraphics[width = 0.8 \linewidth]{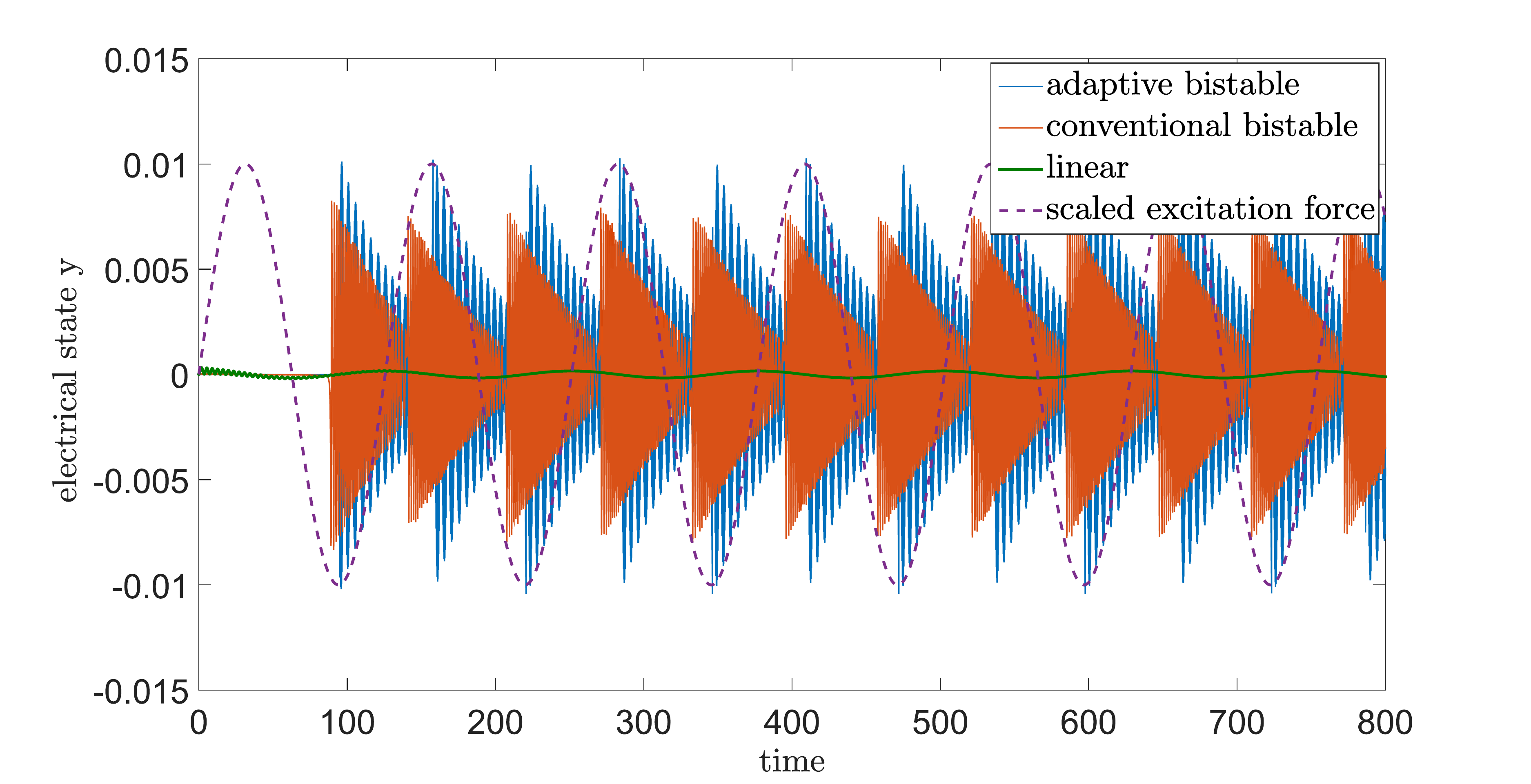}
 \caption{Electrical-state (voltage or current depending on transduction mechanism) time histories of linear, conventional bistable, and adaptive bistable energy harvesters subjected to harmonic excitation with excitation amplitude $F_0$=10, and frequency $\omega$=0.05.}
  \label{Fig:6}
\end{figure}

Figures \ref{Fig:5} and \ref{Fig:6} show time histories of the displacement and electrical-domain state (voltage or current for capacitive or inductive transduction mechanisms, respectively) for the three adaptive bistable, conventional bistable and linear harvesters. According to the figures, although they all have the same maximum displacement, the maximum induced voltage (current) in them is quite different with the adaptive bistable having the largest and the linear having the smallest induced voltage (current). One could also notice the BLSH logic in the adaptive bistable harvester by comparing the moments of the transition from one end to the other and the excitation force extremums. It should also be noted that the conventional bistable harvester is trying to mimic the BLSH strategy in a less effective way.  

\begin{figure}
 \centering
 \includegraphics[width = 0.8 \linewidth]{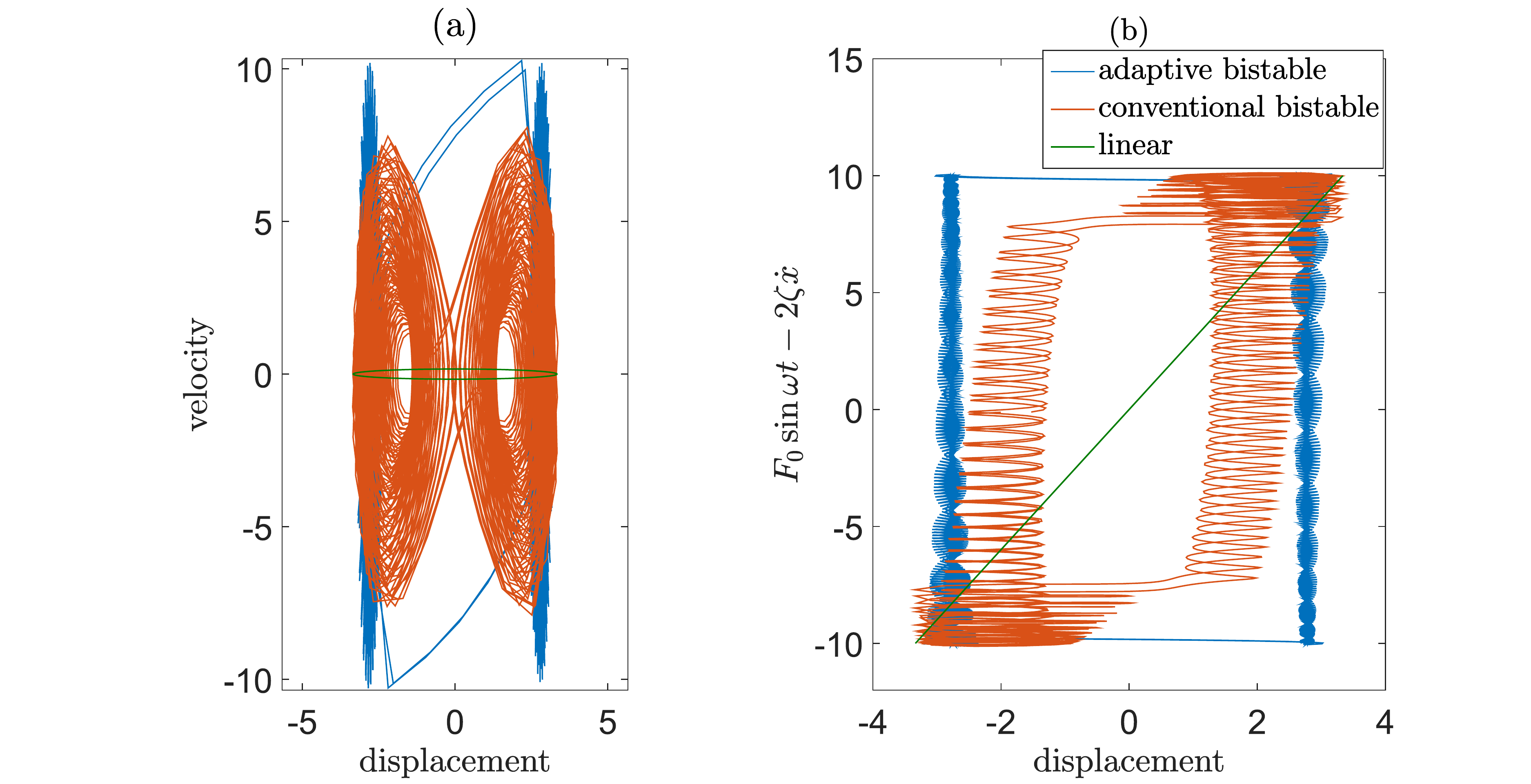}
 \caption{Phase portrait (a), and displacement-force diagram (b) of the three harvesters when subjected to harmonic excitation with excitation amplitude $F_0$=10, and frequency $\omega$=0.05.}
  \label{Fig:7}
\end{figure}

\begin{figure}
 \centering
 \includegraphics[width = 0.8 \linewidth]{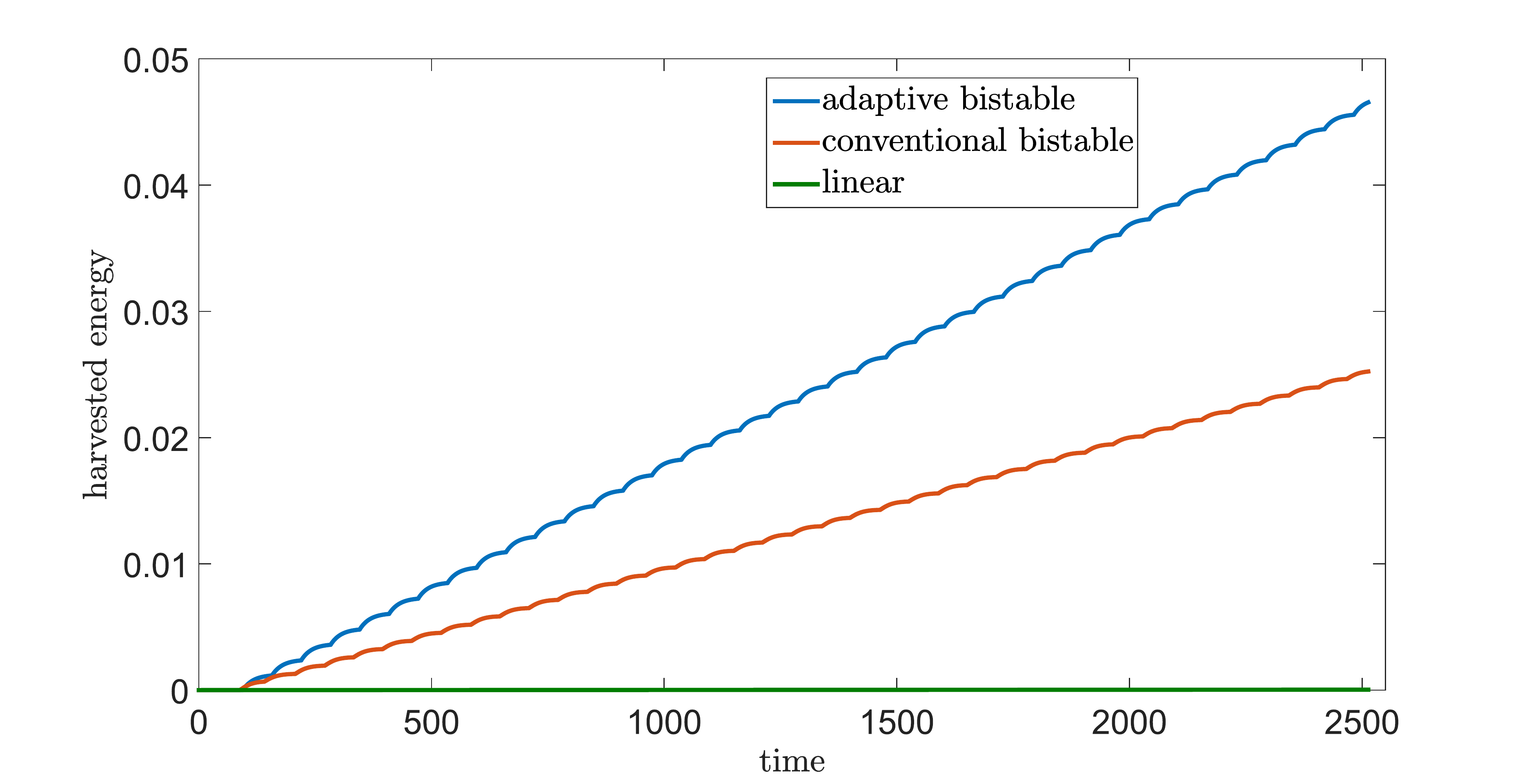}
 \caption{Time history of the harvested energy by the three harvesters when subjected to harmonic excitation with excitation amplitude $F_0$=10, and frequency $\omega$=0.05.}
  \label{Fig:8}
\end{figure}

Another way to compare the harvesters' performances is via their phase portraits. Fig.\ref{Fig:7}(a) depicts these phase portraits. As seen in the figure, the transition of the oscillator's mass between the two displacement limits occur at a higher velocity for the adaptive bistable harvester than the other two. The force-displacement diagram in Fig.\ref{Fig:7}(b) illustrates it even better as how the adaptive bistable harvester outperforms the other two. This diagram shows the force capable of doing positive work versus displacement. An ideal harvester i.e. a harvester with BLSH strategy and ideal harvesting force, will have a perfect rectangle on this diagram, given the displacement limits. This rectangle represents the maximum amount of energy that could be pumped into the harvester (which will be consequently harvested by the ideal harvesting force) in one cycle. The ideal harvester with the perfect rectangle in the force-displacement diagram is very analogous to the Carnot cycle with its perfect rectangle in the temperature-entropy diagram given the temperature limits of the hot and cold reservoirs. In both cases, all the other systems (harvesters and heat engines) fall within this perfect rectangle enclosing a smaller area. Time histories of the harvested energy via the three harvesters depicted in Fig.\ref{Fig:8} prove the higher efficiency of the adaptive bistable system over the other two.

\subsection{random excitation: walking motion}
\begin{figure}
 \centering
 \includegraphics[width =0.9 \linewidth]{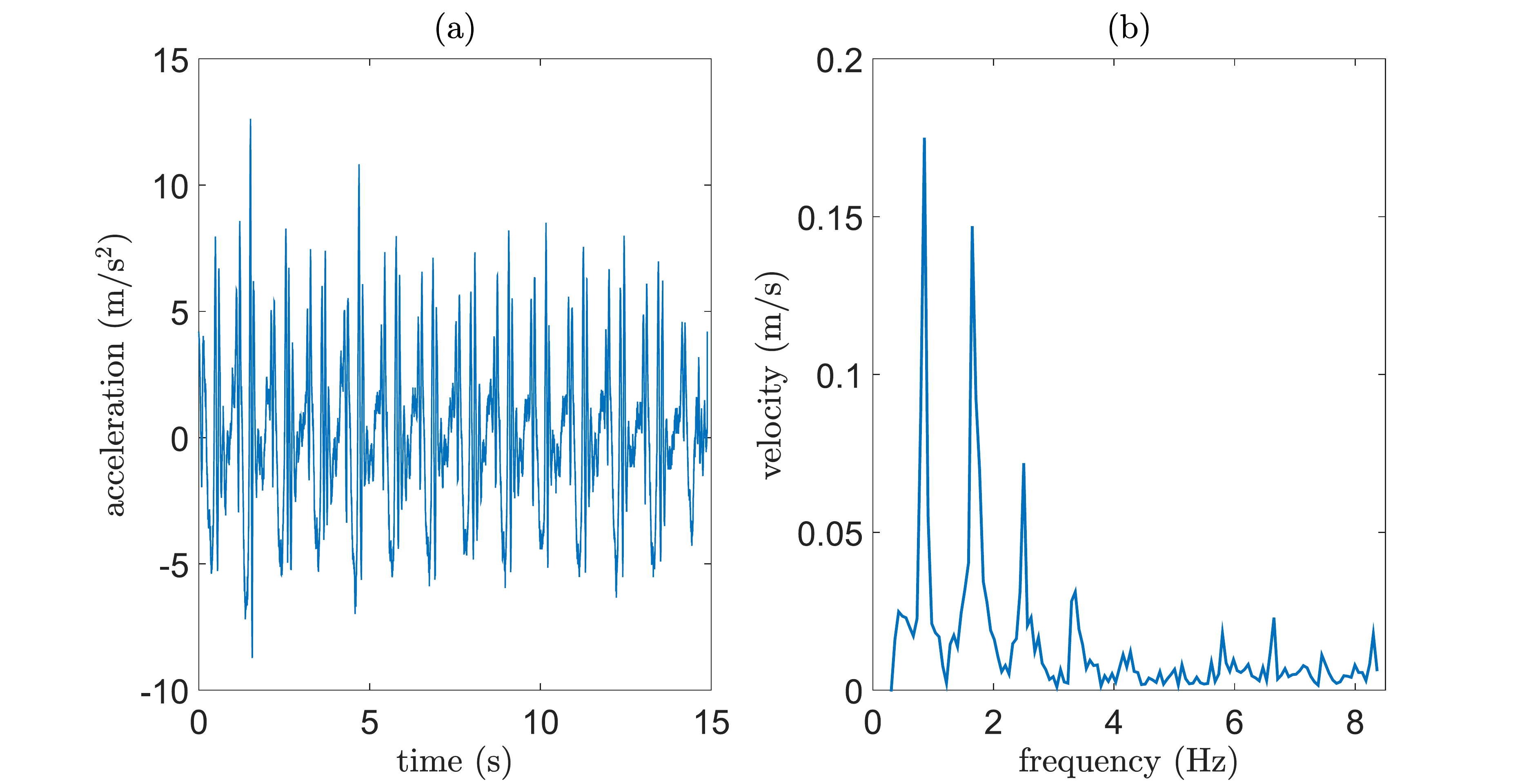}
 \caption{Non-stationary random walking excitation \cite{kluger2014nonlinear}: (a) acceleration time history recorded at the hip while walking, and (b) velocity spectrum (Fourier transform) of the walking motion}
  \label{Fig:9}
\end{figure}

As mentioned earlier, most of the real-world excitations are random and non-stationary rather than harmonic, and that the linear and bistable harvesters do not work efficiently when subjected to these types of excitations. To examine and compare the performance of the three harvesters to random excitations, we subject all the harvesters to experimental and relatively low-frequency walking motion. This data is experimentally recorded at the hip level while walking \cite{kluger2014nonlinear}. The time history and spectral representation of the walking excitation used here are depicted in Fig.\ref{Fig:9}.

For simulations the experimental data is first non-dimensionalized with scaling frequency of 500Hz, and scaling length of 20$\mu \mathrm{m}$. Again, first the conventional bistable potential parameters ($a$, and $x_s$) are optimized for maximum harvested energy for a displacement constraint of 1.5; then the parameters of the adaptive bistable and linear harvesters are set such that they do not exceed this displacement limit. The harvested energy is computed the same way as in the case of the harmonic excitation with the only difference that it is multiplied by the constant $\alpha \kappa^2$ for the sake of easier numerical comparison between different harvesters.

\begin{figure}
 \centering
 \includegraphics[width = 0.9 \linewidth]{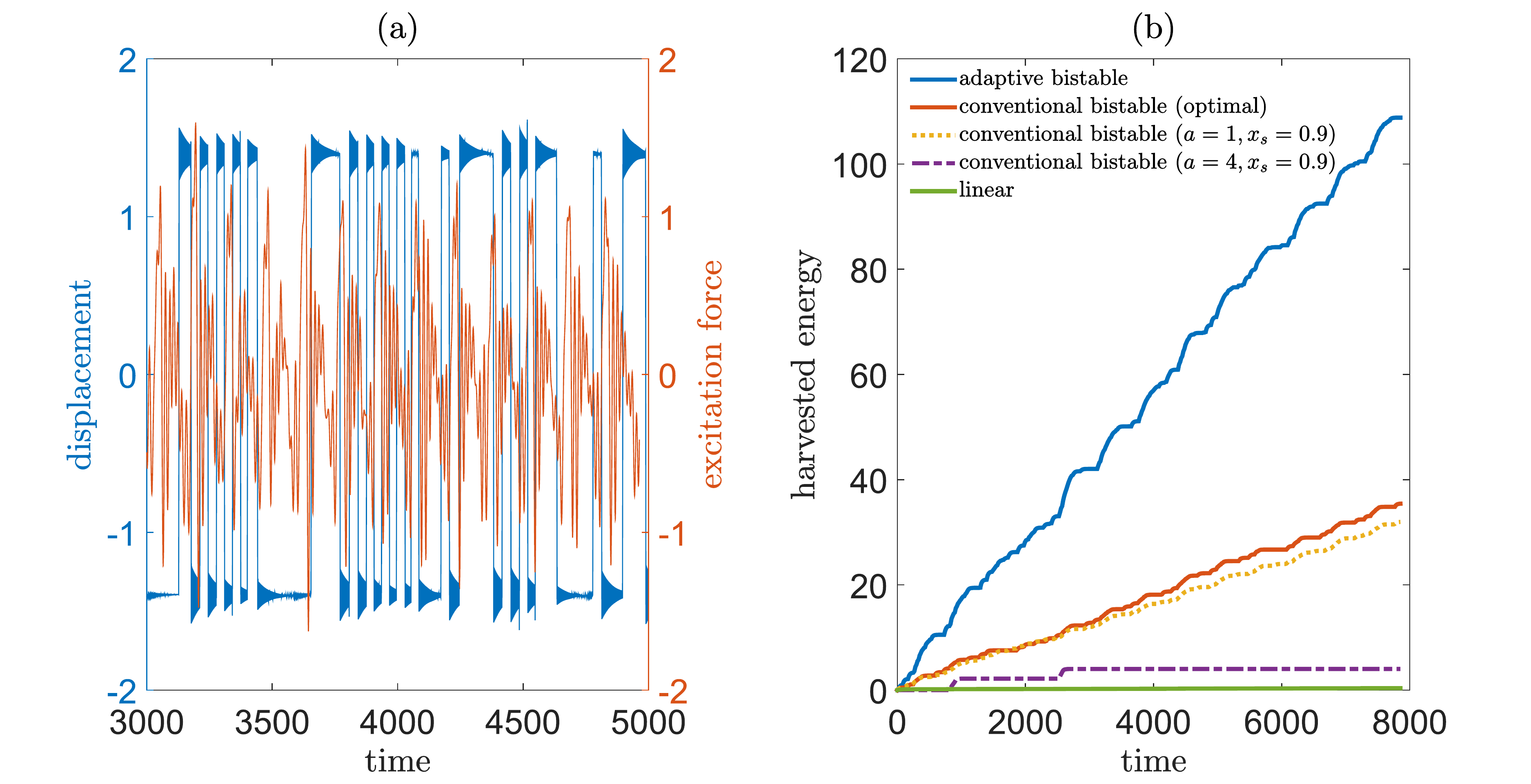}
 \caption{Energy harvesting from walking motion: (a) displacement time history of the harvester mass with adaptive bistability subjected to displacement constraint of $|x_{\mathrm{max}}|<1.5$ (b) energy harvesting time histories of the linear, adaptive bistable, and conventional bistable harvesters. Three conventional bistable harvesters with different parameters are tested.}
  \label{Fig:10}
\end{figure}

Fig.\ref{Fig:10}.(a) illustrates the displacement time history of the harvester with adaptive bistability following a BLSH logic. Harvested energy via the harvesters are compared in Fig.\ref{Fig:10}.(b). In addition to the optimal conventional bistable system ($x_s=0.9$, and $a=1.6$), two other bistable systems with detuned $a$ parameter are also simulated. According to the figure, BLSH adaptive bistable harvester outperforms the optimal conventional bistable and the linear harvesters. It could also be seen that changes in the bistable system parameters could significantly diminish the harvester's efficacy.   

\section{Conclusions}
In this paper, the major drawbacks of the linear and bistable vibration energy harvesters were pointed out and a novel adaptive bistable harvester was proposed to overcome them. The adaptive bistable harvester follows a buy-low-sell-high strategy. In this strategy, the harvester mass is held at the lowest displacement limit ($-x_{\mathrm{max}}$) before the excitation force reaches its maximum which then the harvester mass is pushed to the highest displacement limit ($x_{\mathrm{max}}$) and is held there waiting for the excitation force minimum to reach and do the same thing in the reverse fashion. Although this strategy guarantees maximum harvested energy in an ideal harvester with no mechanical damping, it was shown by simulations that it also works pretty well with more realistic set-ups.

It was also discussed how the adaptive bistable system could be used to enforce the BLSH strategy, and how this could be implemented in practice. It was shown that a harvester equipped with adaptive bistability following a BLSH logic outperforms its linear and conventional bistable counterparts significantly under both harmonic and experimental non-stationary random walking excitations. Also the proposed harvester does not suffer from the robustness issues similar to those of the linear and conventional bistable systems when the system parameters are detuned. Additionally, it was observed that at low-frequency excitations the conventional bistable system tries to mimic the BLSH strategy which gives an insight to why this harvester is more efficient than its linear counterpart at low frequency excitations.

\section*{References}
\bibliography{SMSManuscript.bbl}
\end{document}